\documentclass[aps,twocolumn,amsmath,amssymb,longbibliography]{revtex4-2} 

\usepackage[utf8]{inputenc}
\usepackage[english]{babel}
\usepackage{csquotes}
\usepackage{comment}

\usepackage{tikz, calc, float,xcolor}
\usepackage{amsmath,amsthm,amsfonts,amssymb,braket}
\allowdisplaybreaks

\definecolor{color1}{RGB}{0,80,155}

\newcommand{\dd}[1]{\ensuremath{\mathrm{d}} #1}	
\newcommand{\ii}{\ensuremath{\mathrm{i}}}

\newcommand{\rme}{\ensuremath{\mathrm{e}}}
\newcommand{\D}{\ensuremath{\mathrm{D}}}
\newcommand{\T}{\ensuremath{\mathrm{T}}}
\newcommand{\f}{\ensuremath{\mathrm{F}}}
\newcommand\Wtilde{\stackrel{\sim}{\smash{\mathcal{W}}\rule{0pt}{1.3ex}}_{m,n}}
\newcommand{\F}{\ensuremath{\mathcal{F}}}
\DeclareMathOperator{\I}{I}
\DeclareMathOperator{\II}{II}

\usepackage{hyperref}

\DeclareMathOperator{\arcsinh}{arcsinh}

\DeclareMathOperator{\tr}{Tr}

\begin{document}

	\title{Particle detectors in superposition in de Sitter spacetime}
	\author{Laura Niermann}
	\email{laura.niermann@itp.uni-hannover.de}
	\author{Luis C.\ Barbado}
	\email{luis.cortes.barbado@univie.ac.at}
	\date{\today}

	\begin{abstract}
		Cosmological particle creation is the phenomenon by which the expansion of spacetime results in the production of particles of a given quantum field in that spacetime. In this paper, we study this phenomenon by considering a multi-level quantum particle detector in de Sitter spacetime coupled to a massless real quantum scalar field. Rather than considering a fixed classical trajectory for the detector, following recent novel approaches we consider a quantum superposition of trajectories, in particular of static trajectories which keep a fixed distance from one another. The main novel result is that, due to the quantum nature of the superposition of trajectories, the state of the detector after interaction with the field is not only a mixture of the thermal states that would be expected from each individual static trajectory but rather exhibits additional coherences due to interferences between the different trajectories. We study these in detail and associate them with the properties of the particle absorbed by the detector from the thermal bath.
	\end{abstract}
	
	\maketitle
	\section{Introduction}
	
	One of the best-known results in quantum field theory in curved spacetime is \emph{cosmological particle creation}. It is the effect by which, in an expanding spacetime that initially contains no particles of a given field, particles will be created due to the cosmological expansion. This phenomenon was initially discussed by Parker in \cite{Parker1967, Parker1968, Parker1969} and Sexl and Urbantke in \cite{Sexl1969}. As of today, it is part of the standard literature of the field, such as in \cite{Birrell1982}.
	A cosmological spacetime where it is especially interesting to study particle creation is de Sitter spacetime~\cite{Mottola1985}. De Sitter spacetime is of special interest because, according to the current cosmological paradigm built from the observations of the past decades, it is asymptotically identical to the universe we live in \cite{Schmidt1998}. 
	
	Particle creation in de Sitter spacetime bears close similarities to the Unruh effect in flat spacetime, mainly because de Sitter consists of a constantly accelerated expansion. In general, even free-falling observers in de Sitter will detect the cosmological particle creation taking place in the spacetime, with a temperature proportional to the acceleration rate of the expansion, with an analogous formula to the case of the Unruh effect and proper acceleration. Moreover, local observers in de Sitter spacetime, which remain static with respect to one another, experience proper acceleration that counteracts the spacetime expansion. This proper acceleration increases the temperature of the perceived radiation for these observers in a way that combines the cosmological particle creation and the Unruh effect. 
	The Unruh effect was first introduced in flat spacetime in \cite{Unruh1976}, which is closely related to the works \cite{Davies1976, Fulling1973}. Early studies that considered the Unruh effect in the context of de Sitter spacetime and assigned thermal properties to the de Sitter geometry are \cite{Narnhofer1996, Deser1997, Deser1998, Deser1999}.
	
	In the study of quantum field theory in curved spacetime, it is a well-established practice to consider particle detectors as a way of probing the particle content of a field as perceived by different observers \cite{DeWitt1980, Birrell1982}. These detectors are localized systems with internal degrees of freedom that couple to the field so that reading out the state of these degrees of freedom provides information about the particle perception by the observers following the trajectory of the detector.
	
	Customarily, particle detectors are considered to follow well-defined classical trajectories. In recent works, this situation has been generalized to the case where a particle detector follows a quantum superposition of trajectories, studying phenomena such as the Unruh effect~\cite{Barbado2020} or Hawking radiation~\cite{Paczos2023}. Quantum superposition of trajectories was considered in different contexts \cite{Foo2020, Wood2021, Gale2022, Foo2022, Foo2023, Foo2023a}. In \cite{Zych2020} the authors consider the superposition of detector trajectories in de Sitter spacetime for different scenarios: the superposition of spatially translated trajectory in one de Sitter geometry is considered, and analogies with the superposition of de Sitter spacetimes with different curvatures are drawn. 
	The detector was considered to be an Unruh-de Witt detector with two energy levels as introduced in \cite{DeWitt1980}. The conclusions about the particle perception were discussed in terms of the response function, which, under certain conditions, can be considered as providing the particle detection rate of a collection of detectors.
	
	In the present paper, we generalize the study in \cite{Zych2020} by considering the more general model of detector introduced in \cite{Barbado2020}. This consists of a multi-level particle detector, which allows for an analysis beyond the response function in terms of coherences left between the particle detection along the different trajectories of the quantum superposition. As in \cite{Zych2020}, we consider trajectories that are uniquely distinguishable and keep a fixed distance from one another. The main result we obtain is that, in general, the interaction of the detector with the field is not simply an incoherent mixture of the excitations along the different trajectories in superposition, but rather, some coherences between the trajectories are left. This result is analogous to that found in the context of the Unruh effect \cite{Barbado2020} and Hawking radiation \cite{Paczos2023}. The coherences obtained can be physically discussed as providing information about the spatial profile of the absorbed particles of the field.
	
	The article is structured as follows. In Section \ref{statementProblem}, we describe the setup of the problem, introducing the quantization of the field, the detector, and its trajectories. In Section \ref{calcFinalState}, we compute the excitation of the detector due to the interaction with the field, which constitutes the main result of the work. We discuss the physical interpretation of this result in Section \ref{phys-Interpretation}. We finally close with an outlook of the work in Section \ref{superposCurvature}.

	\section{Statement of the problem} \label{statementProblem}
	
\begin{figure}[h]
		\centering
		\includegraphics[page=1,width=9cm]{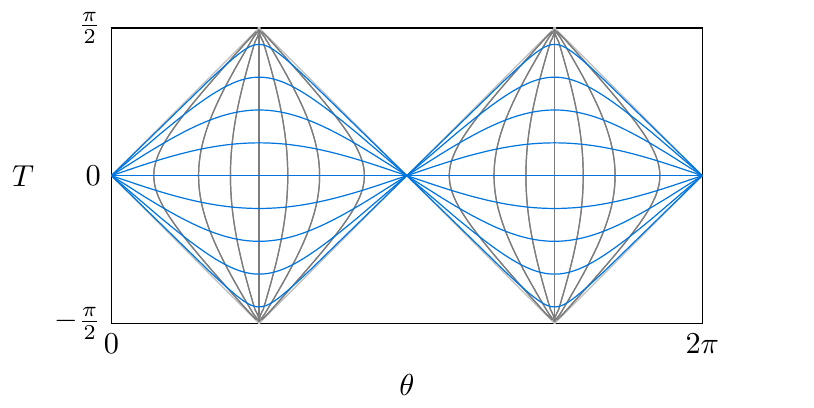}
		\caption{Depiction of static coordinates where blue lines are constant time slices and grey lines trajectories with constant spatial coordinate ($\theta$ is the angle of the unwrapped hyperboloid and $T$ the compactified time related to the coordinate time $t$ with $\cosh t = 1/\cos T$)}
		\label{fig:staticCoords}
	\end{figure}

	Let us consider a family of trajectories in de Sitter spacetime, corresponding to observers who remain static (with fixed proper distance) with respect to one another. We use the parametrization of the spacetime given by static coordinates, which embed $2$ dimensional de Sitter spacetime in $2+1$ dimensional Minkowski space as follows: 
	\begin{align}
		\vec{x}(r,t) =& \begin{pmatrix}	x_0 \\ x_1 \\ x_2	\end{pmatrix}  = \begin{pmatrix}	\sqrt{\ell^2-r^2} \sinh(t/\ell) \\ \pm \sqrt{\ell^2-r^2} \cosh(t/\ell) \\ r	\end{pmatrix}. \label{eq:coordStatic}
	\end{align} Here, $\ell$ is the de Sitter radius, which is directly related to de Sitter's curvature $R=2/\ell^2$ and $t$, and $r$ are the temporal and spatial coordinates (we consider natural units $\hbar = c = 1$). These coordinates satisfy the hyperboloid condition:
	\begin{align}
		-x_0^2 + x_1^2 + x_2^2 = 
		\ell^2
	\end{align} 
	One static patch of de Sitter is described by the static coordinates for $r^2\leq \ell^2$. It corresponds to the area causally accessible to the observers. The induced de Sitter metric in a given static patch, in static coordinates, is then \begin{align}
		g_{\mu\nu} =& \begin{pmatrix}	-1+\frac{r^2}{\ell^2} & 0 \\ 0 & \frac{\ell^2}{\ell^2-r^2}	\end{pmatrix} \label{eq:metricStatic}
	\end{align}
	In Fig.~\ref{fig:staticCoords}, we plot a time-compactified version of the two static patches of de Sitter spacetime.
	
	We consider a massless scalar field $\hat\phi(x)$ as the radiation field, which is defined globally in de Sitter spacetime. We consider the Euclidean vacuum state \cite{Bousso2001}, which is also sometimes referred to as the Bunch-Davies vacuum state \cite[Eq.~(88)]{Higuchi2018}, as the initial state of the field. In \cite[section 6]{Bousso2001}, it is shown that this vacuum state can be expressed as a linear combination of modes in the two static patches, $\I$ and $\II$, as follows:
	\begin{align}
		\ket0_\f =& \prod_{\omega=0}^{\infty} \sqrt{1-e^{-2\pi\omega\ell}} e^{e^{-\pi\omega\ell} (\hat{a}_{\omega}^{\I})^\dagger (\hat{a}_{\omega}^{\II})^\dagger} \ket{\Omega_{\I}} \otimes \ket{\Omega_{\II}} \label{eq:thermal}
	\end{align}
	where $\ket{\Omega_{\I}}$ and $\ket{\Omega_{\II}}$ are the vacuum states of the static patches $\I$ and $\II$ (as perceived by static observers in the respective patches), and $(\hat{a}_{\omega}^{\I})^\dagger$ and $(\hat{a}_{\omega}^{\II})^\dagger$ the creation operators of the respective Fock quantization. It is clear from Eq.~(\ref{eq:thermal}) that the Euclidean vacuum state is perceived as a thermal bath by local observers in the static patches, which reproduces the thermal cosmological particle creation due to the accelerated expansion.
	
	We consider the Unruh-DeWitt model for the detector \cite{DeWitt1980} with two modifications: We consider a multi-level detector, with more than two internal energy levels $\{\ket{0}_{\D},\ \ket{\omega_1}_{\D},\ \ket{\omega_2}_{\D},\ldots\}$ with energies $0 < \omega_1 < \omega_2 < \ldots$; and we consider quantum superpositions of static trajectories in a given static patch. Each well-defined static trajectory corresponds to a state of the set $\{\ket{1}_{\T}, \ket{2}_{\T}, \ket{3}_{\T},\ldots \}$. We consider these trajectories to be fully distinguishable from one another so that this set forms an orthonormal basis of the Hilbert space corresponding to the external degrees of freedom of the detector.
	
	The static trajectories in each path follow the timelike Killing field of that patch, given by~$\partial_t$. These trajectories are:
	\begin{align}
		x_r(\tau) = \begin{pmatrix}
			t(\tau) \\ r(\tau) 
		\end{pmatrix} = \begin{pmatrix}
			\tau / \sqrt{1-r^2/\ell^2} \\ r \label{eq:trajectory}
		\end{pmatrix}
	\end{align}
	Since the relation between the coordinates and the proper time is dependent on the trajectory, we shall elevate the coordinates to an operator acting on the states $\ket{n}_\T$ of the trajectories in the following way:
	\begin{align}
		\hat{x}_r(\tau) \ket{n}_\T = \left(	\tau / \sqrt{1-r_n^2/\ell^2} , r_n\right)\ket{n}_\T. \label{eq:traject}
	\end{align}
	
	We work in the interaction picture, where the detector is coupled to the field with the following interaction Hamiltonian:
	\begin{equation}
		\hat{H}_{\mathrm{I}}(\tau) = \varepsilon \chi (\tau) \hat{m} (\tau) \hat{\phi} (\hat{x}_r(\tau)).
		\label{coupling}
	\end{equation}
	Here $\varepsilon \ll 1$ is a small parameter that controls the intensity of the interaction, $\hat{m} (\tau)$ the detector monopole, and $\chi(\tau)$ a switching function that switches the coupling on and off in time $T$. The monopole moment $\hat{m}(\tau)$ evolves freely according to:
	\begin{equation}
		\hat{m} (\tau) = \sum_i \zeta_i\ \rme^{\ii \omega_i \tau} \ket{\omega_i} \bra{0}_{\D} + \mathrm{h.c.}
		\label{monopole}
	\end{equation}
	Where $\zeta_i$ characterizes de degree of coupling of the different energy levels.
	
	We consider a Gaussian switching function with interaction time $T$:
	\begin{equation}
		\chi (\tau) = \frac{1}{(2 \pi)^{1/4}} \rme^{-\tau^2 / (4 T^2)},
		\label{switching}
	\end{equation}
	We also impose an adiabaticity condition by considering large enough interaction times so that the switching process itself does not introduce spurious transitions in the detector. This leads to the following condition between the interaction time $T$ and the energies $\omega_i$:
	\begin{align}
		T \sim \frac{1}{\varepsilon\omega_1} \gg \frac{1}{\omega_1} \geq \frac{1}{\omega_i} \label{eq:switching-adiab-cond}
	\end{align}

	We prepare the detector in the ground state~$\ket{0}_\D$ and in a general quantum superposition of static trajectories, while the field is in the Euclidean vacuum state~$\ket{0}_F$. The initial state in the asymptotic past is therefore:
	\begin{equation}
		\ket{\Psi (\tau \to -\infty)} = \ket{0}_{\D} \ket{0}_{\f} \left( \sum_n A_n \ket{n}_{\T} \right),
		\label{initial_state}
	\end{equation}
	where the $A_n$ are the amplitudes for the different trajectories in superposition.

	With this setup, the full state at late times to first order in $\varepsilon$ reads:
	\begin{widetext}
		
		\begin{align}
			\ket{\Psi (\tau \to \infty)} =& \left( \hat{\mathrm{I}} + \ii \varepsilon \int_{-\infty}^{\infty}\dd \tau \hat{H}_{\mathrm{I}}(\tau) \right) \ket{\Psi (\tau \rightarrow -\infty)}
			\nonumber\\
			=& \ket{0}_{\D} \ket{0}_{\f} \left( \sum_n A_n \ket{n}_{\T} \right)+ \ii \varepsilon \int_{-\infty}^{\infty}\dd \tau  \chi (\tau) \hat{m} (\tau) \hat{\phi} (\hat{x}_r(\tau)) \ket{0}_{\D} \ket{0}_{\f} \left( \sum_n A_n \ket{n}_{\T} \right).	\label{eq:final_state}
		\end{align}
		
	\end{widetext}

	\begin{figure*}[t]
		\centering
		\begin{tikzpicture}
			\node (a) at (0,6.) {};
			\node (b) at (5.5,6.) {};
			\node (c) at (11,6.) {};
			\node (d) at (0,0) {};
			\node (e) at (5.5,0) {};
			\node (f) at (11,0) {};
			\node (g) at (3,-6.) {};
			\node[below left] at (a) {\includegraphics[height=5cm,page=2]{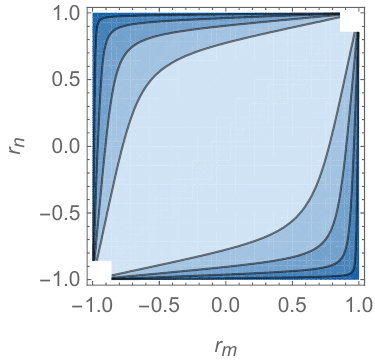}};
			\node[below left] at (b) {\includegraphics[height=5cm,page=3]{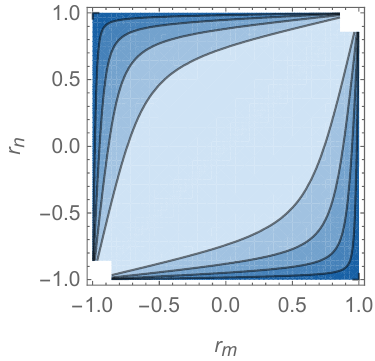}};
			\node[below left] at (c) {\includegraphics[height=5cm,page=4]{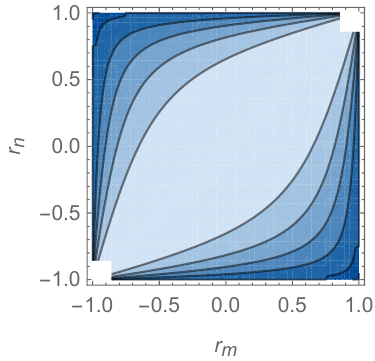}};
			\node[below left] at (d) {\includegraphics[height=5cm,page=5]{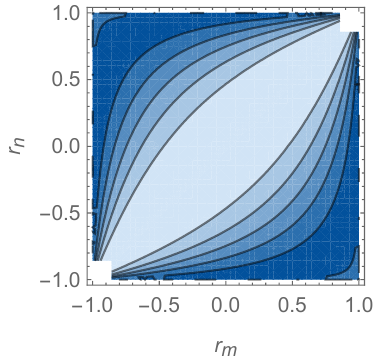}};
			\node[below left] at (e) {\includegraphics[height=5cm,page=6]{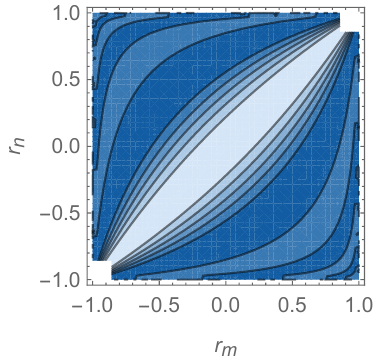}};
			\node[below left] at (f) {\includegraphics[height=5cm,page=7]{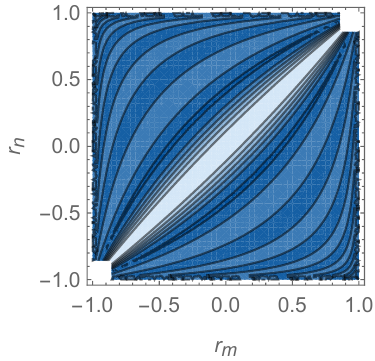}};
			\node[below] at (g) {\includegraphics[width=5cm,page=8]{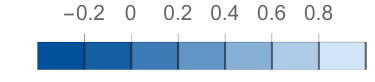}};
			\node[above left] at (a) {$q_{in}=0$};
			\node[above left] at (b) {$q_{in}=0.5$};
			\node[above left] at (c) {$q_{in}=1$};
			\node[above left] at (d) {$q_{in}=2$};
			\node[above left] at (e) {$q_{in}=5$};
			\node[above left] at (f) {$q_{in}=10$};
		\end{tikzpicture}
		\caption{The normalized inner product $\Lambda_{nm}^{ij}$ for different values of $q_{in}$ plotted for the radial coordinates $r_m$ and $r_n$}
		\label{fig:Lambda}
	\end{figure*}
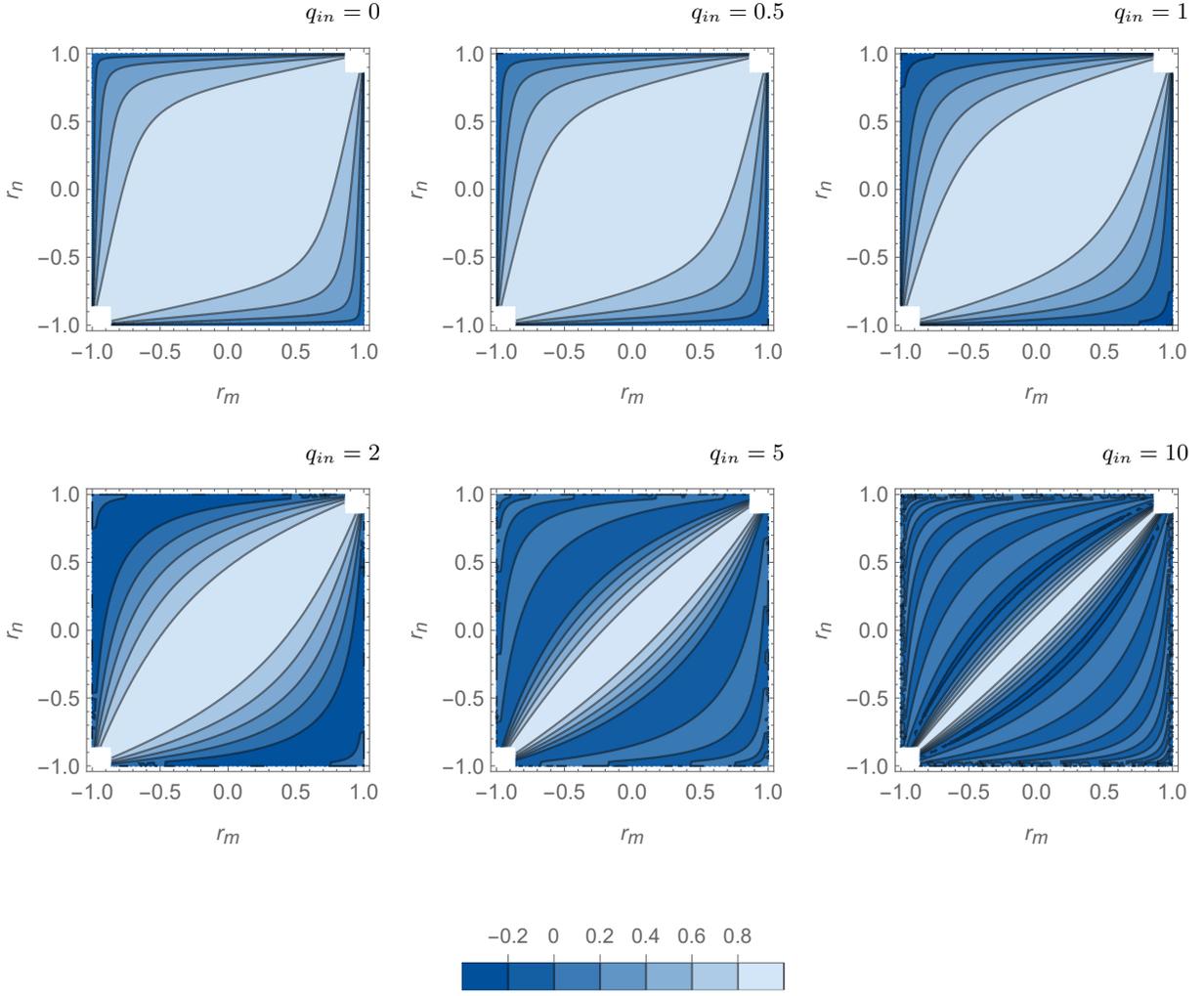

	\section{Final state after the interaction} \label{calcFinalState}
	In order to explicitly compute the final state of the detector, let us re-write the state in~Eq.~(\ref{eq:final_state}) as:
	\begin{align}
		\ket{\Psi (\tau \to \infty)} 
		=& \ket{0}_{\D} \ket{0}_{\f} \left( \sum_n A_n \ket{n}_{\T} \right) \nonumber\\& + \ii \varepsilon \sum_{i,n}   \zeta_i  A_n  \ket{\omega_i}_\D \ket{\omega_i,n}_\f \ket{n}_{\T} \\
		\text{with }\quad \ket{\omega_i,n}_\f =& \int_{-\infty}^{\infty}\dd \tau \chi(\tau) \rme^{\ii \omega_i \tau} \hat{\phi} (\hat{x}_r(\tau))  \ket{0}_{\f} 
	\end{align}
	where we have defined the state $\ket{\omega_i, n}_\f$ as the state in which the field is left when the detector follows the trajectory $\ket{n}_\T$ and gets excited to the state $\ket{\omega_i}_\D$.
	\begin{align}
		\ket{\omega_i,n}_\f =& \int_{-\infty}^{\infty}\dd \tau \chi(\tau) \rme^{\ii \omega_i \tau} \hat{\phi} (\hat{x}_r(\tau))  \ket{0}_{\f} \label{eq:fieldState} \\
		=& \frac{1}{\ii\varepsilon \zeta_i A_n} \bra{\omega_i}_\D \bra{n}_\T \ket{\psi(\tau\rightarrow\infty)} \nonumber \\
		=& \frac{1}{ \zeta_i} \bra{\omega_i}_\D \bra{n}_\T   \int_{-\infty}^{\infty}\dd \tau \hat{H}_{\mathrm{I}}(\tau) \ket{0}_{\D} \ket{0}_{\f}  \ket{n}_{\T} \nonumber
	\end{align}
	
	The density matrix describing the final state of the detector is obtained by tracing out the degrees of freedom of the field:
	\begin{widetext}
		\begin{align}
			\rho_{\D \T} =& \tr_\f\left(\ket{\psi(\tau\rightarrow\infty)} \bra{\psi(\tau\rightarrow\infty)} \right) \nonumber\\
			=& \left( \sum_{m,n} A_m^\ast A_n \ket{n}\bra{m}_\T \right)\ket{0}\bra{0}_\D   +  \varepsilon^2 \sum_{i,j,m,n}  \zeta_j^\ast \zeta_i A_m^\ast A_n  \ket{\omega_i}\bra{\omega_j}_\D \braket{\omega_j,m|\omega_i,n}_\f \ket{n}\bra{m}_{\T}  \label{eq:finalState}
		\end{align}
		In Appendix \ref{app:integral} we calculate in detail the scalar product $\braket{\omega_j,m|\omega_i,n}_\f$. The result yields:
		\begin{align}
			\rho_{\D \T} =&  \left( \sum_{m,n} A_m^\ast A_n \ket{n}\bra{m}_\T \right)\ket{0}\bra{0}_\D  
			+  \varepsilon^2 \frac{ T}{2\pi} \sum_{j,n}  |\zeta_j|^2 |A_n|^2  \ket{\omega_j}\bra{\omega_j}_\D \frac{\omega_j}{e^{2\pi q_{jn}}-1} \ket{n}\bra{n}_{\T} \nonumber\\&
			+  \varepsilon^2\frac{ T}{2\pi}  \sum\limits_{\substack{i,j\\i\neq j}} \sum^{\text{cond}}_{\substack{m,n\\m\neq n}} \zeta_j^\ast \zeta_i A_m^\ast A_n  \ket{\omega_i}\bra{\omega_j}_\D \Lambda_{nm}^{ij}  \frac{\sqrt{\omega_i \omega_j}}{e^{2\pi q_{jn} }-1} \ket{n}\bra{m}_{\T} \label{eq:finalState}
		\end{align}
		
	\end{widetext}

	Let us note that this is an expression up to first order in $\varepsilon$ since $T \sim 1/\varepsilon$ in
	Eq.~(\ref{eq:switching-adiab-cond}). The auxiliary parameter $q_{jn}$ is defined as the ratio:
	\begin{align}
		q_{jn}=\frac{\omega_j}{\kappa_n}, \quad \kappa_n := 1/\sqrt{\ell^2 - r^2}. \label{eq:qdef}
	\end{align}
	The label~`cond' in the sum implies that only the terms for which the following condition holds contribute to the sum:
	\begin{align}
		q_{jn} \approx q_{im}.  \label{eq:q-id}
	\end{align} 
	Finally, the factor $\Lambda_{ij}^{mn}$ corresponds to the normalized inner product between states of the field, given by (as derived in Appendix \ref{app:normInnerProd}):
	\begin{align}
		\Lambda_{ij}^{mn}
		:&= \frac{\braket{\omega_j,m|\omega_i,n}_\f}{\braket{\omega_j,m|\omega_j,m}_\f \braket{\omega_i,n|\omega_i,n}_\f} \nonumber\\
		&= \frac{ \sqrt{\kappa_m \kappa_n} \sin \left(2 q_{in}  \arcsinh\left(\sqrt{b_{m n}}\right)\right)}{\sqrt{2}q_{in}\sqrt{\kappa_n^2+\kappa_m^2} \sqrt{b_{m n} (b_{m n}+1)}} \label{eq:normalizedInnerProduct}\\
		\text{with}\nonumber\\
		b_{mn} :&= \frac{1}{2} \left(\frac{ (1-x_m x_n)}{\sqrt{1- x_m^2} \sqrt{1-
				x_n^2}}-1\right) \quad x_n := r_n/\ell. \label{eq:lambda_value}
	\end{align}
	We plot the quantity~$\Lambda_{ij}^{mn}$ as a function of~$r_n$ and~$r_m$ for different values of~$q_{in}$ in Fig.~\ref{fig:Lambda}.

	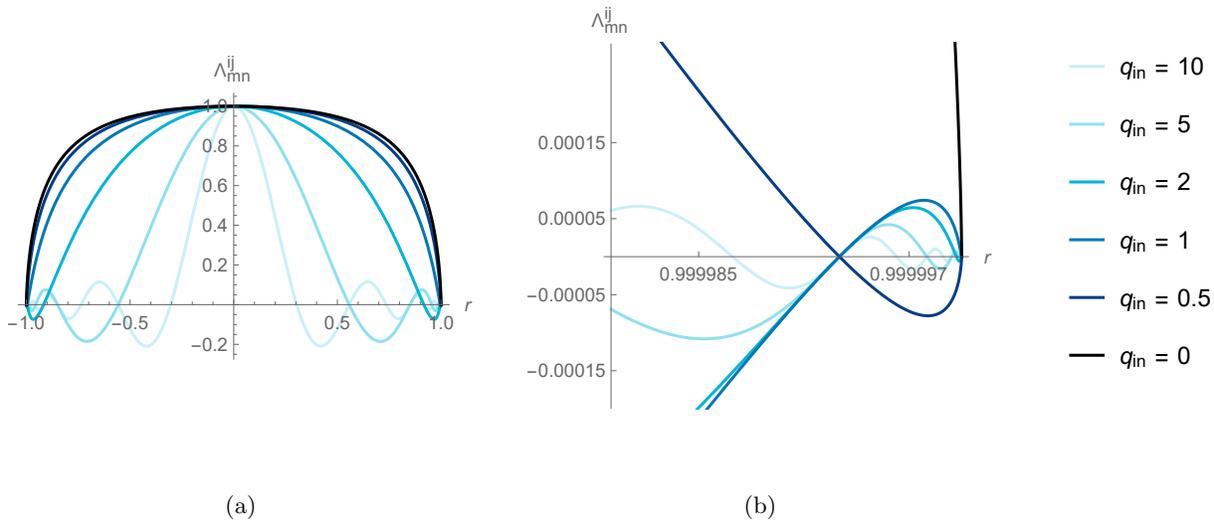
\begin{figure*}
		\centering\begin{tikzpicture}
			\node (a) at (0,0) {};
			\node (b) at (.5,0) {};
			\node (c) at (7.5,0) {};
			\node[left,label={[label distance=5mm]270:(a)}] at (a) {\includegraphics[scale=.7,page=9]{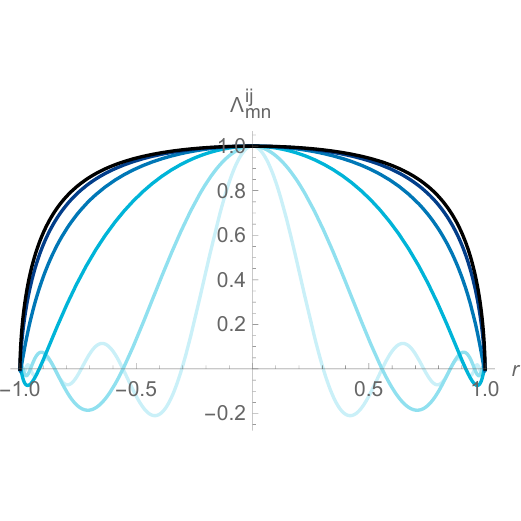}};
			\node[right,label={[label distance=5mm]270:(b)}] at (b) {\includegraphics[scale=.7,page=10]{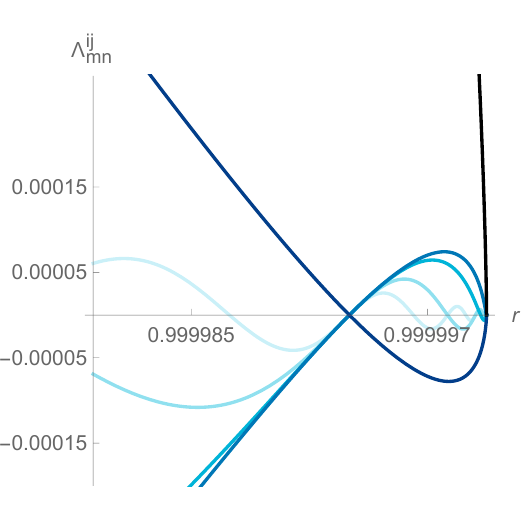}};
			\node[right] at (c) {\includegraphics[scale=.7,page=11]{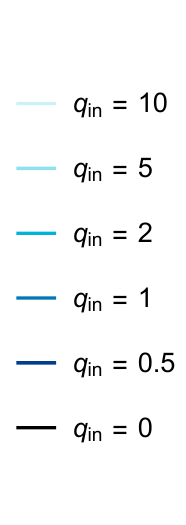}};
		\end{tikzpicture}
		\caption{Values for the normalized inner product $\Lambda_{nm}^{ij}$ where the radial coordinate of one trajectory is fixed to $r_n=0$. This restriction allows us to visualize the oscillations of its value close to the boundaries, which increase for larger values of $q$. (a) shows the entire range of the spatial variable $r$ of the second trajectory, and (b) only shows the area very close to one boundary. The case $q=0$ is the only one without oscillations.}
		\label{fig:oszillations}
	\end{figure*}
	
	If we further want to eliminate the degrees of freedom corresponding to the trajectory of the detector, a possibility would be to trace them. However, considering~Eq.~(\ref{eq:finalState}), one realizes that this completely cancels out any coherences (off-diagonal terms). A way to keep these coherences is to measure the final state of the trajectory in a basis that does not correspond to well-defined static trajectories. Consider that we do so, finding the trajectory to be in the state $\ket{\eta}_\T:=\sum_n B_n \ket{n}_T$. Then the state of the internal energy level up to first order in $\varepsilon$ is given by:
	\begin{widetext}
		\begin{align}
			\rho_\D^\text{measure} = \tr_\T(\ket{\tilde n}\bra{\tilde n}_\T \rho_{\D\T}) =& 
			\left(\sum_{m,n} B_m^\ast A_n^\ast B_n A_m^\ast\right) \ket{0}\bra{0}_\D 
			+   \varepsilon^2 \frac{ T}{2\pi} \sum_{j,n}  |\zeta_j|^2 |A_n|^2 |B_n|^2  \ket{\omega_j}\bra{\omega_j}_\D \frac{\omega_j}{e^{2\pi q_{jn}}-1} \nonumber\\&
			+  \varepsilon^2\frac{T}{2\pi}  \sum\limits_{\substack{i,j\\i\neq j}} \sum\limits_{\substack{m,n\\m\neq n}} \zeta_j^\ast \zeta_i A_m^\ast B_n^\ast A_n B_m   \Lambda_{nm}^{ij}  \frac{\sqrt{\omega_i \omega_j}}{e^{2\pi q_{jn} }-1} \ket{\omega_i}\bra{\omega_j}_\D. \label{eq:main}
		\end{align}
	\end{widetext} 
	
	This is the final state of the detector, which is the main result of the work.

	\section{Physical interpretation} \label{phys-Interpretation}
	
	Let us discuss the result Eq.~(\ref{eq:main}) physically. The zeroth order in $\varepsilon$ corresponds to the contribution of the initial state of the detector, that is, the case where no particles are detected.
	In the first-order perturbation term, we clearly distinguish the diagonal and off-diagonal terms. The diagonal terms correspond to a mixture of Planckian spectrums with the temperatures of the expected perceived radiation along the different static trajectories within the superposition, accordingly weighted depending on the amplitudes and coupling strengths. Therefore, they correspond to the known result of thermal particle detection, with the temperature correspondingly shifted by the \emph{Tolman factor} $\kappa_n$ dependent on the trajectory (defined in Eq.~\ref{eq:traject}), as appearing in Eq.~(\ref{eq:qdef}). Our results are, therefore, consistent with the fact that a detector following a static trajectory in a de Sitter static patch in the Euclidean vacuum perceives a thermal spectrum of particles.
	
	The off-diagonal terms correspond to the coherences, and are the physical novel result. They contain the product of the square roots of the two Planckian spectrums as evaluated in the two energies for which coherences are found. They are also subject to the condition Eq.~(\ref{eq:q-id}) and weighted with the normalized inner product $\Lambda_{nm}^{ij}$ in Eq.~(\ref{eq:normalizedInnerProduct}).
	
	The condition Eq.~(\ref{eq:q-id}) implies that the ratio between the energy of the excitation and the Tolman factor need to be similar (to order $\varepsilon$) for different trajectories in order for coherences to apply. From a physical perspective, this is to be expected since in order for coherences to remain, the excitations on the field along different trajectories need to be not fully distinguishable (the field cannot acquire complete which-path information). Therefore, the energy of the absorbed particle, as described by any observer, has to be similar had it been absorbed along one trajectory or another.
	
	The normalized scalar product for different values of $q_{in}$ is depicted in Fig.~\ref{fig:Lambda}. It takes its maximal value of $1$ if the trajectories coincide, which implies that the term is actually diagonal, and from there, it decays as the trajectories are more separate from one another. The case $q_{in}=0$ is included as depicting the limiting behavior for small frequencies. The oscillatory behavior that superposes to the decay, which is already evident in Fig.~\ref{fig:Lambda}, can be seen better in Fig.~\ref{fig:oszillations} where the normalized scalar product is plotted with one trajectory is fixed to $r=0$. One clearly sees that the oscillations are more significant for higher energies. The only trajectory not showing any oscillatory behavior is the one corresponding to $q_{in}=0$.
	
	We can physically interpret the normalized scalar product as providing a notion of the spatial profile of the absorbed particle from the thermal bath in static coordinates. In fact, we can apply the same reasoning as for the condition Eq.~(\ref{eq:q-id}): Coherences will appear only if the states of the field left along different trajectories are not fully distinguishable, and it is the normalized scalar product that actually measures the indistinguishability. The fact that the scalar product is significant while the trajectories are close and decay as they separate implies that the absorbed particles are somehow delocalized around the static position of the trajectories. One notices also that the decay in the distance is sharper for higher energies, from which we can deduce some qualitative dispersion relation for the delocalized particle.

	\section{Conclusion and outlook} \label{superposCurvature}
	
	In this article, we have studied the behavior of a multi-level particle detector following a quantum superposition of static trajectories in de Sitter spacetime. The main result of the work is that, after the interaction with the quantum radiation field, the state of the detector, in general, retains a coherent superposition of different energy levels corresponding to different trajectories. This behavior is novel both in contrast to the usual setup where only well-defined classical trajectories are considered, and the setup in \cite{Zych2020} where the authors consider a two-level Unruh-de Witt detector, which does not allow to explore the mentioned coherences.
	
	Analogous results to the ones obtained here have been found in the context of the Unruh effect in flat spacetime \cite{Barbado2020} and Hawking radiation perception in Schwarzschild geometry \cite{Paczos2023}, where also an analogous setup was considered. Therefore, this work completes the application of this setup and the study of the corresponding coherences to arguably the remaining relevant phenomenon in quantum field theory in curved spacetime, namely cosmological particle creation.
	
	A natural extension of the present work, following the analyses in \cite{Zych2020, Kabel2022}, would be to consider a comparison between quantum superpositions of different trajectories in a unique background metric, as in this work, and quantum superpositions of different metrics corresponding to different cosmological expansions.

	\newpage
	\begin{appendix}
		\begin{widetext}
			
			\section{Calculation scalar product of states of the field}\label{app:integral}
			To obtain an explicit expression for the final state from Eq.~(\ref{eq:finalState}), we need to calculate the scalar product $\braket{\omega_j,m|\omega_i,n}_\f$. For this, we calculate the corresponding Wightman function in Appendix  \ref{app:Wightman}. An alternative approach would be to use Bogoliubov transformations, which are not employed in this paper.
			
			\begin{align}
				\braket{\omega_j,m|\omega_i,n}_\f =& \int_{-\infty}^{\infty}\dd \tau \int_{-\infty}^{\infty}\dd \tilde{\tau} \chi(\tau) \chi^\ast(\tilde{\tau}) \rme^{\ii (\omega_i \tau-\omega_j \tilde{\tau})} \underset{W(x_m,x_n)}{\underbrace{\bra{0}\hat{\phi}^\dagger(x_m) \hat{\phi} (x_n)  \ket{0}_{\f}}} \nonumber \\
				=& \int_{-\infty}^{\infty}\dd \tau \int_{-\infty}^{\infty}\dd \tilde{\tau} \frac{1}{(2 \pi)^{1/4}} \rme^{-\tau^2 / (4 T^2)}  \frac{1}{(2 \pi)^{1/4}} \rme^{-\tilde\tau^2 / (4 T^2)} \rme^{\ii (\omega_i \tau-\omega_j \tilde{\tau})} W(x_m,x_n) \nonumber\\
				=& \int_{-\infty}^{\infty}\dd \tau \int_{-\infty}^{\infty}\dd \tilde{\tau} \frac{1}{\sqrt{2\pi}} \exp\left[-(\tau^2+\tilde\tau^2)/(4T^2)+\ii\left(\tau \omega_i - \tilde\tau \omega_j\right)\right] W(x_m,x_n) 
				\label{eq:scalarProduct}
			\end{align}

			The Wightman function $W(x_m,x_n)$ refers to trajectories in a de Sitter static patch in a spacetime with de Sitter radius $\ell$ at constant $r_m$ and $r_n$, respectively. The Wightman function is calculated in Appendix \ref{app:Wightman}. We plug this in and obtain the following expression for Eq. (\ref{eq:scalarProduct}): \begin{align}
				\braket{\omega_j,m|\omega_i,n}_\f =& \int_{-\infty}^{\infty}\dd \tau \int_{-\infty}^{\infty}\dd \tilde{\tau} \frac{1}{\sqrt{2\pi}} \exp\left[-(\tau^2+\tilde\tau^2)/(4T^2)+\ii\left(\tau \omega_i - \tilde\tau \omega_j\right)\right] \nonumber\\
				& \cdot \frac{-1}{16 \pi^2} \frac{\kappa_m \kappa_n}{ \sinh^2 \left[(\kappa_m \tilde{\tau}-\kappa_n \tau)/2\right]+\frac12\left(1+\kappa_m r_m \kappa_n r_n-\kappa_m \kappa_n \ell^2\right)-\ii \varepsilon \kappa_m \kappa_n/4}
			\end{align} where $\kappa_{m,n} = 1/\sqrt{\ell^2-r_{m,n}^2}$. 
			To simplify the time dependence of the Wightman function, we use the expansion of the Wightman function in terms of the Fourier modes calculated in Eq.~(\ref{eq:WightmanExpansionFourier}): 
			
			\begin{align}
				\braket{\omega_j,m|\omega_i,n}_\f =&  -\int_{-\infty}^{\infty}\dd{\tau} \int_{-\infty}^{\infty}\dd{\tilde{\tau}} \chi(\tau) \chi^\ast(\tilde\tau) e^{\ii\left(\tau \omega_i - \tilde\tau \omega_j\right)}  \int_{-\infty}^{\infty} -\frac{\kappa_m \kappa_n}{16\pi^2}\frac{2   \sin \left(2 \lambda  \arcsinh\left(\sqrt{b_{m n}}\right)\right)}{\sqrt{b_{m n} (b_{m n}+1)} \left(e^{2 \pi  \lambda }-1\right)} e^{\ii\lambda s} \dd{\lambda}
			\end{align}
			
			where the parameter $b_{mn}$ defined in Eq.~(\ref{eq:abParameter})  captures the dependence on the trajectories $x_m$ and $x_n$ and $s$ is defined to be  $s=(\kappa_m\tilde\tau-\kappa_n\tau)/2$.
			
			This way, we can separate the time-independent part and solve the time integrals: 
			
			\begin{align}
				\braket{\omega_j,m|\omega_i,n}_\f =&  \int_{-\infty}^{\infty} \frac{\kappa_m \kappa_n}{16\pi^2}\frac{2   \sin \left(2 \lambda  \arcsinh\left(\sqrt{b_{m n}}\right)\right)}{\sqrt{b_{m n} (b_{m n}+1)} \left(e^{2 \pi  \lambda }-1\right)} 
						\int_{-\infty}^{\infty}\dd{\tau} \int_{-\infty}^{\infty}\dd{\tilde{\tau}} \chi(\tau) \chi(\tilde\tau) e^{\ii\left(\tau \omega_i - \tilde\tau \omega_j\right)}   e^{\ii\lambda s}
				\dd{\lambda} \label{eq:WightmanSepTime}
			\end{align}
			
			\subsection*{Time integrals}
			
			Using the expression of the Wightman function from Eq.~(\ref{eq:WightmanSepTime}) we can solve the time integrals
			
			\begin{align}
				\int_{-\infty}^{\infty}\dd{\tau} \int_{-\infty}^{\infty}\dd{\tilde{\tau}} \chi(\tau) \chi(\tilde\tau) e^{\ii\left(\tau \omega_i - \tilde\tau \omega_j\right)}   e^{\ii\lambda s}
				&= \int_{-\infty}^{\infty}\dd \tau  \chi(\tau) e^{\ii\left( \omega_i -  \lambda \kappa_n \right)\tau} \int_{-\infty}^{\infty}\dd \tilde{\tau} \chi(\tilde\tau) e^{\ii\left( -  \omega_j + \lambda\kappa_m \right)\tilde\tau}\\
				&= 2\pi \tilde{\chi}(\omega_i-\lambda \kappa_n) \tilde{\chi}(-\omega_j + \lambda \kappa_m)
			\end{align} where $\tilde\chi(\Omega)$ is the Fourier transform of the switching function which again is a Gaussian. 
			The Fourier transform of the switching function takes the following form:
			\begin{align}
				\tilde\chi(\Omega) = \sqrt[4]{\frac{2}{\pi}} T e^{-T^2 \Omega^2}
			\end{align}
			With this, the time-dependent part overall simplifies to \begin{align}
				\int_{-\infty}^{\infty}\dd{\tau} \int_{-\infty}^{\infty}\dd{\tilde{\tau}} \chi(\tau) \chi(\tilde\tau) e^{\ii\left(\tau \omega_i - \tilde\tau \omega_j\right)}   e^{\ii\lambda s} &= 2 \sqrt{2 \pi } T^2 e^{-T^2 \left(\omega_j-\kappa_m \lambda \right)^2} e^{-T^2 \left(\omega_i-\kappa_n \lambda		\right)^2}\\
				&=  2 \sqrt{2 \pi } T^2 e^{-T^2 \kappa_m^2\left(\frac{\omega_j}{\kappa_m}-\lambda \right)^2} e^{-T^2 \kappa_n^2\left(\frac{\omega_i}{\kappa_n}- \lambda		\right)^2}
			\end{align}
			
			We can simplify this further using the adiabaticity assumption from Eq.~(\ref{eq:switching-adiab-cond}) which requires a large interaction time $T$. As the switching function $\chi(\tau)$ is a Gaussian function with a large interaction time, we know that its Fourier transform $\tilde\chi(\Omega)$ is very sharp. Accordingly, the product of the two Fourier-transformed switching functions only contributes if their peaks (the respective means of the Gaussians) are close. The resulting condition is such that the peaks are close and the Gaussians do not vanish is

			\begin{align}
				\frac{\omega_i}{\kappa_n} \approx \frac{\omega_j}{\kappa_m} \label{eq:q-identical}
			\end{align}
			For this quotient, we introduce a quantity \begin{align}
				q_{in}=\frac{\omega_i}{\kappa_n}.
			\end{align} With this quotient, the time integral and, therefore, the product of Fourier-transformed switching functions takes the following form:
			\begin{align}
				\int_{-\infty}^{\infty}\dd{\tau} \int_{-\infty}^{\infty}\dd{\tilde{\tau}} \chi(\tau) \chi(\tilde\tau) e^{\ii\left(\tau \omega_i - \tilde\tau \omega_j\right)}   e^{\ii\lambda s}&= 2 \sqrt{2 \pi } T^2 e^{-T^2 \kappa_m^2\left(q_{jm}-\lambda \right)^2} e^{-T^2 \kappa_n^2\left(q_{in}- \lambda		\right)^2}
			\end{align}
			When plugging this back into the scalar product from Eq.~(\ref{eq:WightmanSepTime}), we obtain
			\begin{align}
				\braket{\omega_j,m|\omega_i,n}_\f =&  \int_{-\infty}^{\infty} \frac{\kappa_m \kappa_n}{16\pi^2} \frac{2   \sin \left(2 \lambda  \arcsinh\left(\sqrt{b_{m n}}\right)\right)}{\sqrt{b_{m n} (b_{m n}+1)} \left(e^{2 \pi  \lambda }-1\right)}  2 \sqrt{2 \pi } T^2 e^{-T^2 \kappa_m^2\left(q_{jm}-\lambda \right)^2} e^{-T^2 \kappa_n^2\left(q_{in}- \lambda		\right)^2}\dd{\lambda} 
			\end{align}
			
			which can be simplified using $q=q_{in}=q_{jm}$
			
			\begin{align}
				\braket{\omega_j,m|\omega_i,n}_\f =&  \frac{\kappa_m \kappa_n}{16\pi^2} \frac{    4 \sqrt{2 \pi }  T^2}{\sqrt{b_{m n} (b_{m n}+1)}}\int_{-\infty}^{\infty} \frac{\sin \left(2 \lambda  \arcsinh\left(\sqrt{b_{m n}}\right)\right)}{ e^{2 \pi  \lambda }-1}  e^{-T^2 (\kappa_m^2+ \kappa_n^2)\left(q-\lambda \right)^2}\dd{\lambda} \label{eq:WightmanEnergyIntegral}
			\end{align}
			\subsection*{Solve Fourier integral}
			We approximate the $\lambda$-integral from Eq.~(\ref{eq:WightmanEnergyIntegral}) by again using that the interaction time $T$ is large, which was imposed by the condition introduced in (\ref{eq:switching-adiab-cond}) and then solve the integral using Laplace's method:
			\begin{align}
				\int_a^b \dd{x} f(x) e^{- n g(x)} \sim \sqrt{\frac{2\pi}{n g''(x_0)}} f(x_0) e^{-n g(x_0)} \qquad \text{for } n\rightarrow \infty
			\end{align} Here, the function $g(x)$ has to be differentiable twice (with a strict minimum such that $g' (x_0)=0$) and $f(x_0) \neq 0$.
			We identify the different terms of the integrals as follows where our integration variable is $\lambda$:

			\begin{align}
				n =& T^2 \\
				f(\lambda) =& \frac{\sin \left(2 \lambda  \arcsinh\left(\sqrt{b_{m n}}\right)\right)}{ e^{2 \pi  \lambda }-1} \\
				g(\lambda) =&(\kappa_m^2+\kappa_n^2)\left(q-\lambda \right)^2, \qquad  \qquad g''(\lambda) = 2  (\kappa_n^2+\kappa_m^2)\\
				g'(\lambda_0) =& -2(\kappa_m^2+\kappa_n^2)(q-\lambda_0) = 0 \quad \Rightarrow \quad \lambda_0 = q_{in}
			\end{align}
			
			For large $T$, the following holds:

			\begin{align}
				\braket{\omega_j,m|\omega_i,n}_\f 
				=& \frac{\kappa_m \kappa_n}{16\pi^2} \frac{4 \sqrt{2} \pi  T  }{\sqrt{b_{m n} (b_{m n}+1)}} \frac{1}{\sqrt{\kappa_n^2+\kappa_m^2}}\frac{ \sin \left(2 q  \arcsinh\left(\sqrt{b_{m n}}\right)\right)}{ e^{{2 \pi  q} }-1} \label{eq:WightmanEnergyIntegralLaplace}
			\end{align}		
			
			\section{Wightman function} \label{app:Wightman}
			
			For a massless scalar field, the positive frequency Wightman function is (see \cite[Eq.~3.59]{Birrell1982})
			\begin{align}
				D^+(x,x') = -\frac{1}{4\pi^2}\frac{1}{(x^0-(x')^0-\ii\varepsilon)^2 - |\vec{x}-\vec{x}'|^2} 
			\end{align} where $\vec{x}$ refers to the spatial part.
			
			In static coordinates evaluated at $x=x_m$ and $x'=x_n$ we can expand this in the components up to first order in $\varepsilon$:
			
			\begin{align}
				-(x_m^0-x_n^0-\ii\varepsilon)^2 + |\vec{x}_m-\vec{x}_n|^2=&-(x^0_m -x^0_n - \ii \varepsilon)^2 + (x^1_m-x^1_n)^2 +(x^2_m-x^2_n)^2  \nonumber\\
				=&-\left(\sqrt{\ell^2-r_m^2} \sinh \left(\frac{t_m}{\ell}\right)-\sqrt{\ell^2-r_n^2} \sinh
				\left(\frac{t_n}{\ell}\right) - \ii\varepsilon\right)^2 \nonumber\\&+\left(\sqrt{\ell^2-r_m^2} \cosh \left(\frac{t_m}{\ell}\right)-\sqrt{\ell^2-r_n^2} \cosh
				\left(\frac{t_n}{\ell}\right)\right)^2+(r_m-r_n)^2\\
				=&-\left(\sqrt{\ell^2-r_m^2} \sinh \left(\frac{t_m}{\ell}-\frac{\ii\varepsilon}{2}\right)-\sqrt{\ell^2-r_n^2} \sinh
				\left(\frac{t_n}{\ell}+\frac{\ii\varepsilon}{2}\right) \right)^2 \nonumber\\&+\left(\sqrt{\ell^2-r_m^2} \cosh \left(\frac{t_m}{\ell} - \frac{\ii\varepsilon}{2}\right)-\sqrt{\ell^2-r_n^2} \cosh
				\left(\frac{t_n}{\ell} + \frac{\ii\varepsilon}{2}\right)\right)^2+(r_m-r_n)^2\nonumber\\
				=&- 2 \sqrt{\ell^2-r_m^2}\sqrt{\ell^2-r_n^2} \cosh\left(\frac{t_m}{\ell}-\frac{t_n}{\ell} - \ii\varepsilon \right) + 2\ell^2 -2 r_m r_n 
			\end{align}
			
			we replace the coordinate time $t$ with the proper time $\tau_i=\sqrt{1-r_i^2/\ell^2} t_i$ and 
			introduce the parameter $\kappa_i = \frac{1}{\sqrt{\ell^2-r_i^2}}$ which simplifies the expression as follows (with $\tau_m=\tilde\tau$ and $\tau_n=\tau$): \begin{align}
				-(x_m^0-x_n^0-\ii\varepsilon)^2 + |\vec{x}_m-\vec{x}_n|^2
				=& 2\ell^2 - \frac{2  \cosh\left(\kappa_m \tau - \kappa_n \tau - \ii\varepsilon \right)}{\kappa_m \kappa_n} + r_m r_n\\
				=& -\frac{2 \left(\cosh (\kappa_m \tilde{\tau}-\kappa_n \tau - \ii\varepsilon)+\kappa_m r_m \kappa_n r_n-\kappa_m \kappa_n \ell^2\right)}{\kappa_m \kappa_n}\\
				=& -\frac{2}{\kappa_m \kappa_n} \left(\cosh (\kappa_m \tilde{\tau}-\kappa_n \tau - \ii\varepsilon)+\sqrt{\kappa_m^2 \ell^2-1} \sqrt{\kappa_n^2 \ell^2-1}-\kappa_m \kappa_n \ell^2\right)
			\end{align}
			
			For the Wightman function, it follows 
			\begin{align}
				W(x_m,x_n) =& -\frac{1}{4\pi^2}\frac{1}{(x^0-(x')^0-\ii\varepsilon)^2 - |\vec{x}-\vec{x}'|^2} \nonumber\\
				=& -\frac{\kappa_m \kappa_n}{8\pi^2}\frac{1}{\cosh (\kappa_m \tilde{\tau}-\kappa_n \tau - \ii\varepsilon)+\sqrt{\kappa_m^2 \ell^2-1} \sqrt{\kappa_n^2 \ell^2-1}-\kappa_m \kappa_n \ell^2} 
			\end{align}
			We use the hyperbolic identity \begin{align}
				\sinh^2\left(\frac{x}{2}\right) = \frac12 \left(\cosh(x)-1\right) \qquad \Leftrightarrow \qquad \cosh(x) = 1+2\sinh\left(\frac{x}{2}\right)
			\end{align}
			The Wightman function is 
			\begin{align}
				W_{m,n}(s) =& -\frac{\kappa_m \kappa_n}{8\pi^2}\frac{1}{1+2\sinh^2 \left(\frac{\kappa_m \tilde{\tau}-\kappa_n \tau}{2} - \ii\varepsilon\right)+\sqrt{\kappa_m^2 \ell^2-1} \sqrt{\kappa_n^2 \ell^2-1}-\kappa_m \kappa_n \ell^2} \\
				=&  -\frac{\kappa_m \kappa_n}{16\pi^2} \frac{1}{\sinh^2\left((\kappa_m \tilde{\tau}-\kappa_n \tau)/2-\ii\varepsilon\right)-b_{mn}} \label{eq:Wightman-s}
			\end{align}with $\kappa_i = 1/\sqrt{\ell^2-r_i^2}$ and 
			\begin{align}
				b_{mn} =& -\frac12\left(1+\kappa_m r_m \kappa_n r_n-\kappa_m \kappa_n \ell^2\right) 
				= \frac{1}{2} \left(\kappa_m \kappa_n
				l^2-\sqrt{\kappa_m^2 \ell^2-1} \sqrt{\kappa_n^2 \ell^2-1}-1\right).
				\label{eq:abParameter}
			\end{align} As we know, that $r_i^2\leq \ell^2$. We can directly see that $b_{mn}$ is non-negative when introducing a rescaled variable $x_i = \frac{r_i}{\ell}$ whose absolute value is always smaller than one. For the parameter $b_{mn}$ we obtain\begin{align}
				b_{mn} = \frac{1}{2} \left(\frac{ (1-x_m x_n)}{\sqrt{1- x_m^2} \sqrt{1-
						x_n^2}}-1\right)
			\end{align}Here we can see that $b_{mn}$ vanishes for identical trajectories ($x_m=x_n$) and is positive otherwise.\\
			
			\subsection*{Expansion in Fourier modes}
			Here, we derive the Fourier expansion of the Wightman function in terms of its variable 
			\begin{align}
				s = \kappa_m \tilde{\tau}-\kappa_n\tau
			\end{align}The variable $s$ is variable whose dimension is the product from time and acceleration. The Wightman function from Eq.~(\ref{eq:Wightman-s}), which we now expand in terms of its Fourier modes, is \begin{align}
				W_{m,n}(s) = \frac{a_{mn}}{\sinh^2(s /2-\ii\varepsilon)-b_{m n}} \hspace{2cm}\text{with}\quad a_{mn} = -\frac{\kappa_m \kappa_n}{16\pi^2} \label{eq:Wightman-t}
			\end{align}
			
			The Fourier transform of the Wightman function is defined in terms of the variable $\lambda$
			\begin{align}
				\Wtilde(\lambda) =\F\left[W_{m,n}(s)\right](\lambda) = \frac{1}{\sqrt{2\pi}}	\int_{-\infty}^{\infty} W_{m,n}(s) e^{-\ii\lambda s} \dd{s} \label{eq:WightmanInt}
			\end{align}
			
			The poles of the Wightman function lie at \begin{align}
				s_n = \left\{ 
				\pm {2 \arcsinh\left(\sqrt{b_{m n}}\right)}+{2 \ii \pi  n} \right\} \quad \text{with }n\in\mathbb{Z}
			\end{align}
			
			We use the Residue theorem to evaluate the integral: \begin{align}
				\oint_\gamma W_{m,n}(s) e^{-\ii\lambda s} \dd{s} = 2\pi\ii \sum_{n=1}^N I(\gamma,s_n) \mathrm{Res}(W_{m,n}(s) e^{-\ii\lambda s},s_n)
			\end{align}
			where $t_n$ are the poles, the winding number $I(\gamma,t_n)$ is one if the pole is in the interior of $\gamma$ and 0 if the pole is outside. Its sign depends on the orientation of the curve: for clockwise integration curves, we get a minus sign. The integration contour $\gamma$ is chosen along the real axis and closed via a half circle in the negative imaginary plane around all poles with negative $n$.
			\begin{align}
				\mathrm{Res}_n^1&=\mathrm{Res}\left(W_{m,n}(s) e^{-\ii\lambda s}, {2 \arcsinh\left(\sqrt{b_{m n}}\right)}+{2 \ii \pi  n} \right) = \frac{a_{mn} e^{2 \lambda  \left(\pi  n-\ii \arcsinh\left(\sqrt{b_{m n}}\right)\right)}}{\sqrt{b_{m n} (b_{m n}+1)}}\\
				\mathrm{Res}_n^2&=\mathrm{Res}\left(W_{m,n}(s) e^{-\ii\lambda s},-{ 2 \arcsinh\left(\sqrt{b_{m n}}\right)}+{2 \ii \pi  n} \right) = -\frac{a_{mn} e^{2 \lambda  \left(\pi  n+\ii \arcsinh\left(\sqrt{b_{m n}}\right)\right)}}{\sqrt{b_{m n} (b_{m n}+1)}}\\
			\end{align}
			As a solution to the integral \ref{eq:WightmanInt}, we get
			\begin{align}
				\oint_\gamma W_{m,n}(s) e^{-\ii\lambda s} \dd{s} =& -2\pi\ii \sum_{n=1}^\infty  \left(\mathrm{Res}_{-n}^1 + \mathrm{Res}_{-n}^2\right)
				= -\frac{4 \pi  a_{mn} \sin \left(2 \lambda  \arcsinh\left(\sqrt{b_{m n}}\right)\right)}{\sqrt{b_{m n} (b_{m n}+1)} \left(e^{2 \pi  \lambda }-1\right)}\\
				\Wtilde(\lambda) =& \frac{1}{\sqrt{2\pi}}\oint_\gamma W_{m,n}(s) e^{-\ii\lambda s} \dd{s}  = -\frac{1}{\sqrt{2\pi}}\frac{4 \pi  a_{mn} \sin \left(2 \lambda  \arcsinh\left(\sqrt{b_{m n}}\right)\right)}{\sqrt{b_{m n} (b_{m n}+1)} \left(e^{2 \pi  \lambda }-1\right)}
			\end{align}
			
			With the inverse Fourier transform, we can express the Wightman function in terms of its Fourier modes: 
			\begin{align}
				W_{m,n}(s) = \frac{1}{\sqrt{2\pi}} \int_{-\infty}^{\infty} \Wtilde(\lambda) e^{\ii\lambda s} \dd{\lambda} = - \int_{-\infty}^{\infty} \frac{2  a_{mn} \sin \left(2 \lambda  \arcsinh\left(\sqrt{b_{m n}}\right)\right)}{\sqrt{b_{m n} (b_{m n}+1)} \left(e^{2 \pi  \lambda }-1\right)}  e^{\ii\lambda s} \dd{\lambda} \label{eq:WightmanExpansionFourier}
			\end{align}

			\section{Diagonal terms of scalar product} \label{app:idTraj}
			
			For obtaining the diagonal terms of the scalar product, we consider two identical trajectories $r=r_m=r_n$ and express $b_{mn}$ in terms of the dimensionless variable $x$, which is the rescaled radial variable:
			\begin{align}
				b_{mn} = \lim_{x_m\rightarrow x_n} \frac{1}{2} \left(\frac{ (1-x_m x_n)}{\sqrt{1- x_m^2} \sqrt{1- x_n^2}}-1\right) = 0
			\end{align}
			
			We consider the result from Eq.~(\ref{eq:WightmanEnergyIntegralLaplace}) and consider the limiting case of identical trajectories:
			\begin{align}
				\braket{\omega_i,m|\omega_i,m} =&  \frac{\kappa_m \kappa_n}{16\pi^2} \frac{4 \sqrt{2} \pi  T  }{e^{{2 \pi  q} }-1} \frac{1}{\sqrt{\kappa_m^2+\kappa_m^2}} \lim_{b\rightarrow 0}  \frac{ \sin \left(2 q  \arcsinh\left(\sqrt{b_{m n}}\right)\right)}{ \sqrt{b_{m n} (b_{m n}+1)}}\\
				=&  \frac{\kappa_m \kappa_n}{16\pi^2}\frac{4 \sqrt{2} \pi  T  }{e^{{2 \pi  q} }-1} \frac{1}{\sqrt{2}\kappa_m} 2q \label{eq:scalar-prod-diag}
			\end{align} 
			Note, that we introduced the parameter $q \approx q_{im}=\frac{\omega_i}{\kappa_m}$. We can plug this back in to obtain
			\begin{align}
				\braket{\omega_i,m|\omega_i,m} 
				=&  \frac{\kappa_m \kappa_n}{16\pi^2}\frac{4\pi  T  }{e^{{ \frac{2 \pi \omega_i}{\kappa_m}} }-1} \frac{1}{\kappa_m} 2\frac{\omega_i}{\kappa_m} 
			\end{align} 
			This is the thermal spectrum with the de Sitter temperature $T_\text{dS}=\frac{\kappa}{2\pi}$: 
			\begin{align}
				\braket{\omega_i,m|\omega_i,m} = \frac{ T}{2\pi}\frac{\omega_i}{\left(e^{{  \omega_i}/{T_\text{dS}} }-1\right)}   = \frac{ T}{2\pi}\frac{\omega_i}{\left(e^{2\pi q }-1\right)}\label{eq:scalar-prod-diag}
			\end{align}

			\section{Normalized inner product} \label{app:normInnerProd}
			Here, we calculate the inner product of normalized states
			\begin{align}
				\Lambda_{nm}^{ij} = \frac{\braket{\omega_j,m|\omega_i,n}_\f}{\sqrt{\braket{\omega_i,n|\omega_i,n}_\f \braket{\omega_j,m|\omega_j,m}_\f}}
			\end{align}
			with this we can expand the off-diagonal terms ($i\neq j$ and $n\neq m$) of the scalar product as follows:
			\begin{align}
				\braket{\omega_j,m|\omega_i,n}_\f =& \Lambda_{nm}^{ij} \sqrt{\braket{\omega_i,n|\omega_i,n}_\f \braket{\omega_j,m|\omega_j,m}_\f} \\
				=& \Lambda_{nm}^{ij}  \frac{ T}{2\pi} \sqrt{\frac{\omega_i \omega_j}{\left(e^{2\pi q }-1\right) \left(e^{2\pi q }-1\right)}}\\
				=& \Lambda_{nm}^{ij}  \frac{ T}{2\pi} \frac{\sqrt{\omega_i \omega_j}}{e^{2\pi q }-1}
			\end{align}
			
			With this, we can see that the information of the off-diagonal terms is encoded in the normalized inner product. For further calculations, we need to plug in the explicit expression of the parameters from Eq.~(\ref{eq:abParameter}) and the results of the scalar product from Eq.~(\ref{eq:WightmanEnergyIntegralLaplace}) and its diagonal terms from Eq.~(\ref{eq:scalar-prod-diag}). We also use, that $q_{in} = \frac{\omega_i}{\kappa_n}$.
			
			\begin{align}
				\Lambda_{ij}^{mn}
				=& \frac{\kappa_m \kappa_n}{2\pi^2} \frac{\sqrt{2}\pi^2    \sin \left(2 q  \arcsinh\left(\sqrt{b_{m n}}\right)\right)}{\sqrt{b_{m n} (b_{m n}+1)}\sqrt{\kappa_n^2+\kappa_m^2}\sqrt{\omega_i\omega_j}}\\
				=& \frac{ \sqrt{\kappa_m \kappa_n} \sin \left(2 q  \arcsinh\left(\sqrt{b_{m n}}\right)\right)}{\sqrt{2}q\sqrt{\kappa_n^2+\kappa_m^2} \sqrt{b_{m n} (b_{m n}+1)}}
			\end{align}
			
			As a sanity check, we look at the normalized inner product for identical trajectories, which is obtained by taking the limit $b_{nn}\rightarrow 0$:
			
			\begin{align}
				\Lambda_{ii}^{nn}=& \frac{ \sqrt{\kappa_n \kappa_n} \sin \left(2 q  \arcsinh\left(\sqrt{b_{n n}}\right)\right)}{\sqrt{2}q\sqrt{\kappa_n^2+\kappa_n^2} \sqrt{b_{n n} (b_{n n}+1)}} 
				=  \frac{ \sqrt{\kappa_n \kappa_n}}{\sqrt{2}q\sqrt{2\kappa_n^2} } \,2q =1
			\end{align}
		\end{widetext}
	\end{appendix}

	\section*{Acknowledgements}

	L.N. acknowledges support by the Quantum Valley Lower Saxony (QVLS), the DFG through SFB 1227 (DQ-mat), the RTG 1991, and funded by the Deutsche Forschungsgemeinschaft (DFG, German Research Foundation) under Germany's Excellence Strategy EXC-2123 QuantumFrontiers 390837967.\\
	L.C.B. acknowledges financial support by the Austrian Science Fund (FWF) through BeyondC (F7103-N48), by the Spanish Government through the project PID2020-118159GBC43/AEI/10.13039/501100011033 and by the Junta de Andaluc\'{\i}a through the project FQM219. This publication was made possible through the support of the ID 61466 grant from the John Templeton Foundation, as part of The Quantum Information Structure of Spacetime (QISS) Project (qiss.fr). The opinions expressed in this publication are those of the authors and do not necessarily reflect the views of the John Templeton Foundation.
	
	\bibliography{bibliography}
	
\end{document}